\documentclass[twocolumn]{aastex631}

\usepackage{graphicx}
\usepackage{ulem}
\usepackage{amsmath}
 
\def\eq#1{\begin{equation}#1\end{equation}}
\def\eqlbl#1{\label{eq:#1}}
\def\eqref#1{(\ref{eq:#1})}

\def\dd{\mathrm d}
\def\sinc{{\rm sinc}} %Fred
\def\nablab{{\mbox{\boldmath $\nabla$}}}
\def\R{{\widetilde R}}

\shorttitle{CONTRACTION BY COOLING OR DESPINNING OF CONDENSED PLANETS}
\shortauthors{Ricard and Chambat}
\graphicspath{{./}{figures/}}
\begin{document}

\title{
MASS-RADIUS RELATIONSHIPS AND CONTRACTION OF CONDENSED PLANETS BY COOLING OR DESPINNING}

\author[0000-0002-0998-4670]{Yanick Ricard}
\affiliation{Ecole normale sup\'erieure, CNRS \\
15 parvis Ren\'e Descartes \\
Lyon, F-69007, France}

\email{Ricard@ens-lyon.fr}

\author[0000-0001-8967-077X]{Fr\'ed\'eric Chambat}
\affiliation{Ecole normale sup\'erieure, CNRS \\
15 parvis Ren\'e Descartes \\
Lyon, F-69007, France}

\email{Frederic.Chambat@ens-lyon.fr}

%\collaboration{6}{(AAS Journals Data Editors)}

%\author{Butler Burton}
%\affiliation{Leiden University}
%\affiliation{AAS Journals Associate Editor-in-Chief}
%\altaffiliation{AASTeX v6+ programmer}
%\affiliation{AAS Director of Publishing}
%\affiliation{American Astronomical Society \\
%1667 K Street NW, Suite 800 \\
%\author{Magaret Donnelly}
%\affiliation{IOP Publishing, Washington, DC 20005}



\begin{abstract}

Condensed planets contract or expand as their temperature changes. With the exception of the effect of phase changes, this phenomenon is generally interpreted as being solely related to the thermal expansivity of the planet's components. However, changes in density affect pressure and gravity and, consequently, the planet's compressibility. A planet's radius is also linked to its rate of rotation. Here again, changes in pressure, gravity and compressibility are coupled. In this article we clarify how the radius of a condensed planet changes with temperature and rotation, using a simple and rigorous thermodynamic model. We consider condensed materials to obey a simple equation of state which generalizes a polytopic EoS as temperature varies. Using this equation, we build simple models of condensed planet's interiors including exoplanets, derive their mass-radius relationships, and study the dependence of their radius with temperature and rotation rate. We show that it depends crucially on the value of $\rho_s g R/K_s$ ($\rho_s$ being surface density, $g$ gravity, $R$ radius, $K_s$ surface incompressibility). This non-dimensional number is also the ratio of the dissipation number which appears in compressible convection and the Grune\"isen mineralogic parameter. While the radius of small planets depends on temperature, this is not the case for large planets with large dissipation numbers; Earth and a super-Earth like CoRoT-7b are in something of an intermediate state, with a moderately temperature-dependent radius. Similarly, while the radius of these two planets are functions of their rotation rates, this is not the case for smaller or larger planets.
\end{abstract}


%%%%%%%%%%%%%%%%%%%%%%%%
\section{Introduction}
%%%%%%%%%%%%%%%%%%%%%%%%

In the 19th century, the contraction of our planet during its secular cooling was sometimes invoked to explain the Earth's topography \citep{dana}. This interpretation was abandoned after the advent of plate tectonic theory, and mountain ranges have since been explained by plate collisions and interactions. However, for other planets and satellites in the solar system which do not appear to 
have plate tectonics and have not undergone a resurfacing event like appears to be the case for Venus, many features such as compression scarps are attributed to a reduction in the planet's radius.
Conversely, the absence of obvious signs of planet compression or extension has been used to put constrains on planet's thermal evolution. The amount of planet radius variation has been discussed in the case on Mercury  \citep[3 to 7 km of contraction, e.g.][]{byrne14}, Mars \citep[0 to 4 km of contraction since the Early Noachian, e.g.][]{nahm-schultz} or the Moon  \citep[negligible radius variation, e.g.][]{solomon-chaiken}. 

The change in radius of a planet can have various origins. One obvious cause of contraction is thermal cooling, i.e. the increase in density of the mineralogical phases when the temperature decreases. Another possible contraction linked to thermal cooling is the phase changes that can occur. For exemple, crystallization of a liquid metallic core to form a denser solid inner core is necessarily accompanied by a decrease in planetary radius. Another cause may be a change in angular velocity. The despinning of a planet 
has two effects. The first reduces the hydrostatic flattening  \citep{chambat} and induces a stress pattern that favors lithospheric cracks oriented parallel to the rotation axis. Such a preferential  orientation of faults has been invoked  on Mercury 
\citep{melosh88}. This early weakening of the lithosphere may later be reactivated by thermal cooling \citep{watters98,byrne14}. The second effect contracts the planet by decreasing the spherically averaged centrifugal force \citep{saito74}.

In a convecting planet, the temperature is controlled by the adiabatic gradient, and cooling  or heating occurs at all depths. A simple relation between a planet's cooling rate and the variation $\delta R$ of its radius $R$ is often used, for example
 \eq{{\delta R\over R}={1\over R^3} \int_0^R \alpha(r) \delta T(r) r^2 \dd r,\eqlbl{simple}}
which takes into account the variations of thermal expansivity $\alpha$ and temperature $T$, with depth \citep{solomon-chaiken}, or even
\eq{{\delta R\over R}={1\over 3} \bar \alpha \overline{\delta T},\eqlbl{simple2}}
where $\bar \alpha$ and $ \overline{\delta T}$ are uniform \citep{hauck04}.
For the Earth, an estimate of the cooling can be obtained from petrologic observation of ancien rocks. For example, the composition of non-arc basaltic rocks as a function of age, suggests a cooling rate of 50-100 K Gyr$^{-1}$ \citep{herzberg10}. The thermal expansivity is decreasing with depth in the mantle but a typical value of $\alpha = 3 \times 10^{-5}$~K$^{-1}$ would imply a radius reduction of 10-20 km in 3 Gyr.

These relations \eqref{simple}-\eqref{simple2} are in fact problematic. Choosing an appropriate thermal expansivity for a quantity that varies with time and depth is a first difficulty. But the real difficulty lies elsewhere: changing the temperature affects the density, the gravity and therefore the pressure. It is well known that for large objects, density is mainly related to pressure, rather than temperature, so that it is not obvious that the change in radius of a cooling planet is so directly related  to temperature. \citet{jaupart15} give an approximated estimate of the pressure change influence by supposing a small and uniform compressibility. Of course, a model that gives us the radial variations of thermal expansion, incompressibility, density and temperature as functions of depth, makes it possible to accurately compute the radius change as a fonction of temperature by perturbing the planet's elasto-gravitational equations but such a model is only known for the Earth. 

To calculate the variations of the Earth's radius with the variations of its angular velocity $\Omega$,  \citet{saito74} has applied this method of perturbing the elasto-gravitational equations to obtain 
 \eq{\frac{\delta R}{R} = \frac{2}{3} h_0^* \frac{\Omega^2 R}{g}  \frac{\delta \Omega}{\Omega},\eqlbl{saito}}
 where $g$ being Earth's surface gravity and $h_0^*$ is a degree0 rotational Love number estimated to be $h_0^*=0.09835$ using a radial model very close to PREM. 
 
Is seems therefore that precise calculation of the contraction of a planet by cooling or despinning is only possible for the Earth. In this paper, we aim to show that a realistic estimate can, however, be obtained using a simple model for a generic condensed planet. 
%%%%%%%%%%%%%%%%%%%%%%%%
\section{A Lane-Emden planet}
%%%%%%%%%%%%%%%%%%%%%%%%
\subsection{Equation of state}
To get an estimate of the relation between the radius and temperature of a cooling planet, we must first choose an equation of state (EoS) to describe the thermodynamic relation between a planet's pressure $P$, temperature $T$ and density $\rho$. For a condensed planet (solid or liquid, silicate of metal), Murnaghan's EoS \citep{murnaghan37} gives a fairly simple and versatile expression that fits well with various high-pressure, high-temperature experiments on silicates and metals, as well as with the radius properties of the Earth \citep{ricard22,ricard23}:
\eq{P={K_0\over p}\left[\left({\rho\over\rho_0}\right)^p-1\right]+\alpha_0 K_0 (T-T_0).\eqlbl{EoS1}}
In this expression $\alpha_0$, $K_0$, $\rho_0$ and $T_0$ are the thermal expansivity, isothermal incompressibility, density and
temperature under reference conditions (e.g. zero pressure and 25$^\circ$C), and $p$ a non-dimensional empirical exponent.
This Eos leads to simple expressions for thermal expansivity and  isothermal incompressibility, which are related only to density
\eq{\alpha=\alpha_0 \left( {\rho_0\over\rho} \right)^p, K_T= K_0 \left({\rho\over\rho_0} \right)^p, \eqlbl{alpha-K}}
and these relations are reasonably verified experimentally \citep{anderson79}. 

 When $T$ is uniform, this equation \eqref{EoS1} belongs to the family of polytropic EoS, on the form 
\eq{P=a {\rho}^{1+{1\over n}}+P_b,\eqlbl{EoS2}}
where $n$ is called the polytropic index (the exponent $p$ in \eqref{EoS1} is therefore $1+1/n$).
Polytropic equations  have been extensively used in astrophysical litterature since \citet{chandrasekhar39}
where the focus is often in stars and gaseous planets, and the index $n$ is large 
 ($n=+\infty$ for a perfect gas). Neutron stars are well modeled with polytropic index close to, but lower than 1.
In a solid planet $p\approx 3-4$ or $n\approx 1/3-1/2$ \citep{stixrude05}. The exponent $p$ is also $\partial K_T/\partial P$
and therefore it quantifies the increase of incompressibility with pressure.

\subsection{Density equation}
If we consider a non-rotating planet   with spherical symmetry (we account for the planet rotation in section 5), its  
density verifies an equation that simply reformulates the gravity equation (Poisson's equation)
\eq{4\pi G\rho= - \nablab \cdot {\mathbf g} = -\nablab \cdot \left({\nablab P\over \rho} \right)= -\nablab \cdot\left(K {\nablab \rho \over \rho^2} \right) \eqlbl{poisson}}
In this expression, the quantity $K$ is defined by
\eq{K=\rho {\dd P\over \dd \rho}.}

If the planet is convecting, its state is close to an adiabatic state with an isentropic incompressibility$K_S$, and $K$ can therefore be identified with $K_S$ \citep{bullen75}. However, for a condensed planet, $K_s$ is very close to $K_T$ and
similarly 
the two heat capacities $C_P$ and $C_V$ are comparable. This is a consequence of the thermodynamic rules (Mayer's relation) that imply $K_S/K_T=C_P/C_V= 1+\Gamma \alpha T$ where  the Gr\"uneisen parameter $\Gamma=\alpha  K_T / \rho C_V $ is of order 1 for solid silicates, liquid metal (and even for ideal gases), and decreases slightly with depth, while $\alpha T$ is a small quantity of order 10$^{-2}$ (in planets, $T$ increases with depth but $\alpha$ decreases more, see equations \eqref{alpha-K}). Whether the planet is convecting or not, the influence of temperature on density remains negligible compared to that of pressure and we can therefore identify $K$ with $K_T$ or $K_S$. Similarly we
confuse the heat capacities $C_P$ and $C_V$ that we later note $C$. 

\subsection{Non dimensional variables}
\begin{figure}[htbp]
\begin{center}
\centerline{\includegraphics[width=9cm,angle=0]{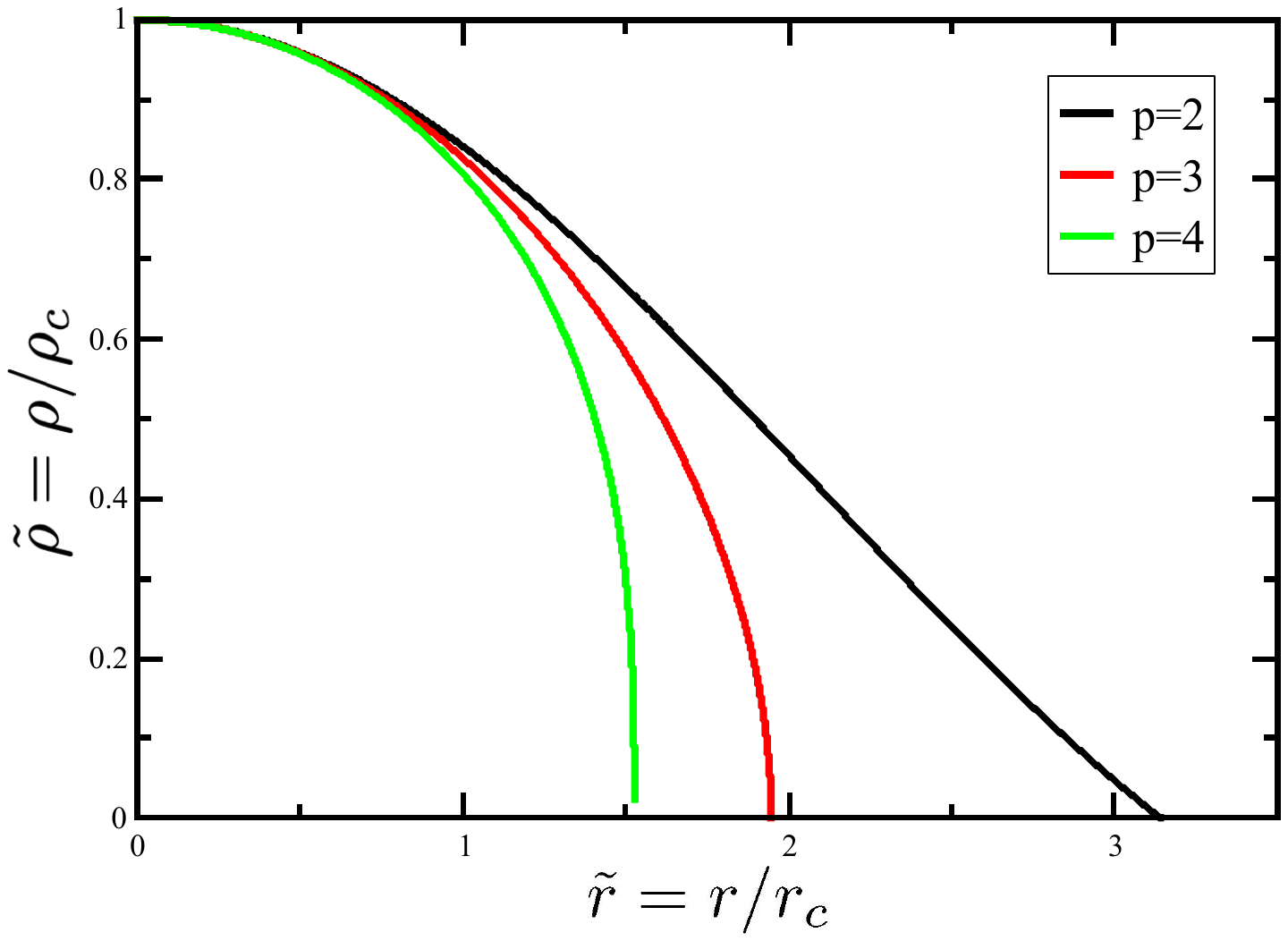}}
\caption{The non-dimensional density profiles $\tilde\rho(r/r_c)$, for $p=2$ (solid black),
$p=3$ (solid red), and $p=4$ (solid green). }
\label{rc}
\end{center}
\end{figure}

We now define a dimensionless density 
$\tilde \rho = \rho/\rho_c $ and a dimensionless radius
$\tilde r = r/r_c $
where $\rho_c$ is the density at the center of the planet
(in the paper, the subscript
$c$ will always refer to the properties at the center, the subscript $s$ at the surface, the subscript $0$ to the reference values of the EoS \eqref{EoS1}, and the tilde symbol
will refer to variables without dimensions)
and $r_c$ is defined by:
\eq{r_c^2=r_0^2 \left( {\rho_c \over \rho_0} \right)^{p-2}  \eqlbl{rtilde}}
where the length $r_0$ is
\eq{r_0^2={K_0\over 4\pi G\rho_0^2}. \eqlbl{defL}}
Using these variables,  Equation \eqref{poisson} becomes
 \eq{ { 1\over \tilde r^2} {\dd\over \dd \tilde r} \left(\tilde r^2 {\tilde\rho^{p-2}} {\dd \tilde\rho \over \dd \tilde r}\right)+\tilde \rho=0 \eqlbl{base}.}
This equation must be solved with $\tilde \rho(0)=1$ and ${\dd \tilde \rho/\dd \tilde r }(0)=0$, until the outer non dimensional radius $\tilde r=\widetilde R$, where
\eq{\widetilde R={R\over r_c}. \eqlbl{Rsrc}}
The resolution of \eqref{base}  is analytical for $p=2$,
in which case the solution is the first arc of a $\sinc$ function, and for $p=\infty$ and $p=6/5$ (i.e. for a polytropic index 0 or 5) which are not in the appropriate range of parameters for a condensed planet. For the other exponents $p$, 
the dimensionless equation \eqref{base} can be easily solved numerically using a Runge-Kutta method. The solutions of \eqref{base} as a function of $\tilde r$ are shown in Figure \ref{rc} for $p=2$, 3 and 4. These curves are computed until $\tilde \rho=0$ but the density with dimensions are only related to these curves for $\tilde r \le \widetilde R$, i.e. for $\tilde \rho(\tilde r)\ge\rho_s/\rho_c$ where $\rho_s$ is the surface density.

Notice that, following  \citet{chandrasekhar39} the density of stellar objects in astrophysical literature is often sought in the form
$\tilde \rho= \theta^n=\theta^{1\over p-1}$ \citep{Horedt2004PolytropesAI}. 
In which case \eqref{base} becomes the well known Lane-Emden equation
\eq{\tilde \nabla^2 \theta + \theta^n=0\eqlbl{L-E}}
when the radius in the Laplace operator $\nabla^2$ is normalized by $r_c/\sqrt{p-1} $. However, solving numerically
the Lane-Emden equation \eqref{L-E} or the equation \eqref{base} presents the same degree of difficulty and we thought that the planetary
physicists might prefer thinking in terms of $\rho$ than $\theta$.

\section{Density with physical dimension}

\subsection{Input parameters}
To obtain a solution with physical dimensions, various quantities need to be specified. We need to know the central density $\rho_c$, the planet radius $R$ and the non-dimensionalizing length $r_c$, but the choice of $r_c$  is related to that of two EoS parameters $K_0$ and $\rho_0$.
Therefore, four quantities need to be specified, $(\rho_c, R, K_0,\rho_0)$. Alternatively we can choose  another set of four independent parameters from which $(\rho_c, R, K_0,\rho_0)$ can be derived. For example, we will calculate the mass of  planets when $(\rho_s, R, K_0,\rho_0)$ are chosen, i.e. when the planet's surface density and radius $R$ are specified. To apply our model to the Earth or to all planets for which mass and radius are known, we will instead choose $(M,R,K_0,\rho_0)$ and calculate the appropriate
values of the central and surface densities. 
Finally, in the paragraph where we discuss the effects of rotation,  $(\rho_s, M,K_0,\rho_0)$
will be fixed values, and $R$ will vary with the rate of rotation.

When $(\rho_s, R, K_0,\rho_0)$ are known, the definition of $r_c=R/\widetilde R$, (see \eqref{rtilde}) using $\rho_c=\rho_s / \tilde \rho(\widetilde R)$ implies an implicit equation for $\widetilde R$
\eq{ \left({R\over \widetilde R}\right)^2={K_0\over 4 \pi G \rho_0^p} \left( {\rho_s\over \tilde \rho (\widetilde R)}\right)^{p-2}.\eqlbl{rhosur}} 
Similarly when $(M, R,K_0,\rho_0)$ are known, we start from
\eq{M=4\pi \int_0^R \rho r^2 \dd r=4\pi \rho_c r_c^3 \int_0^\R \tilde \rho \tilde r^2 \dd \tilde r,
\eqlbl{mass0}} 
which can also be written, using \eqref{base}, as
\begin{equation}
\begin{split}
M&=-4\pi \rho_c r_c^3 \!\int_0^{\R} \! {\dd\over \dd \tilde r} \left(\tilde r^2 {\tilde\rho^{p-2}} {\dd \tilde\rho \over \dd \tilde r}\right) \! \dd\tilde r \\
&= -{4\pi \rho_c R^3  \over \R}  \tilde \rho^{p-2} (\R) {\dd \tilde \rho \over \dd \tilde r} (\R).  \eqlbl{mass}
\end{split}
\end{equation}
To get again an  implicit equation for $\widetilde R$, we can
eliminate $\rho_c$ from this equation using \eqref{rtilde} and \eqref{Rsrc}, to get
\eq{M = -\left( { (4\pi)^{p-1} G \rho_0^p R^{3p-4}   \over K_0 {\widetilde {R^{}}^p}}\right)^{1\over p-2}   \tilde \rho^{p-2}(\R)  {\dd \tilde \rho \over \dd \tilde r} (\R).  \eqlbl{rcKx}}
Finally when $(\rho_s, M, K_0,\rho_0)$ are known, we eliminate
$R$ from \eqref{rcKx} using \eqref{rhosur}  which leads to a third implicit equation of $\widetilde R$.
In each of the three cases, when $\widetilde R$ is determined, the central and surface densities or the mass are readily obtained.

In the remainder of this article, we take $K_0=130$GPa which is a typical incompressibility near the surface of the Earth
and $\rho_0=4000$ kg m$^{-3}$ (for planets made of silicates and metal, we thought reasonable to choose a reference density somewhat larger than that of crustal rocks or very shallow mantle, see also \citep{ricard22}, but this choice is not crucial). The length $r_0$ is therefore $r_0=3113$ km.
At depth this incompressibility increases continuously with the density according to\eqref{alpha-K}.  For exemple, a seismologic model of the Earth like PREM indicates
that mantle density increases with depth by a factor of 1.7 (from $\approx 3200$ kg m$^{-3}$ to $\approx 5400$ kg m$^{-3}$) while incompressibility increases by a
factor 5.0 (to $\approx 650$ GPa). Using \eqref{alpha-K}, this implies $p=\ln 5.0/\ln 1.7=3.03$ which again confirms that $p\approx 3$ is
an appropriate choice. The Earth's density has also several large discontinuities with depth (in the transition zone, the core-mantle and
inner-outer core boundaries) while the incompressibility only exhibits minor discontinuities with depth. This is another argument that led us to prefer, for our continuous model, a choice of numerical values that corresponds to the observation of the relatively continuous behavior of incompressibility. 

Notice that $K_0$ and $\rho_0$  appear in the equations only by the ratio $K_0/\rho_0^p$
in the definition \eqref{rtilde} of $r_c$. Compositionally denser (resp. lighter) materials have often a larger (resp. smaller) incompressibility, for example, $K_0/\rho_0^3$ for silicates and water are similar  (2.03 and 2.1 Pa m kg$^3$, using $K_0=2.1$ GPa and $\rho_0=1000$ kg m$^{-3}$ for water). The exponent $p$ appropriate for water or ices is also in the range of those appropriate for silicates, close to 4 \citep{fei93}. Our model can therefore be extended to a larger variety of compositions than silicate planets.

\subsection{Density profile}

We first apply our model to various planets for which $M$ and $R$ are known, see Table~\ref{Table1}. We consider five planets or satellites of the solar system (Moon, Ganymede, Mars, Mercury and Earth) and also examine the case for the exoplanet CoRot-7b using the mass and radius determinations of \citet{corotM} and \citet{corotR}. The uncertainties 
suggested in these two papers may be underestimated as other articles have proposed values outside the corresponding confidence intervals. Among the planets selected, we include Ganymede whose composition
probably has equal parts of rocky material and water, liquid or ices.

\begin{table}[htp]
\caption{Mass, radius and average densities of various planets. }
\begin{center}
\begin{tabular}{ccccc}
\hline
                &   $M$                &   $\bar\rho$   & $R$ \\
                &   $10^{22}$ kg  & kg m$^{-3}$  & km   \\
\hline
CoRoT-7b & 3620    & 9360 & 9735 \\
Earth         &   597.2 & 5515 & 6371   \\
Mars         &   64.17 & 3933 & 3389   \\
Mercury    &   33.01 & 5427 & 2439    \\
Ganymede & 14.82  & 1940 & 2631 \\
Moon        &   7.342 & 3344 & 1737  \\
\end{tabular}
\end{center}
\label{Table1}
\end{table}

\begin{figure}[h]
\begin{center}
\centerline{\includegraphics[width=9.1cm,angle=0]{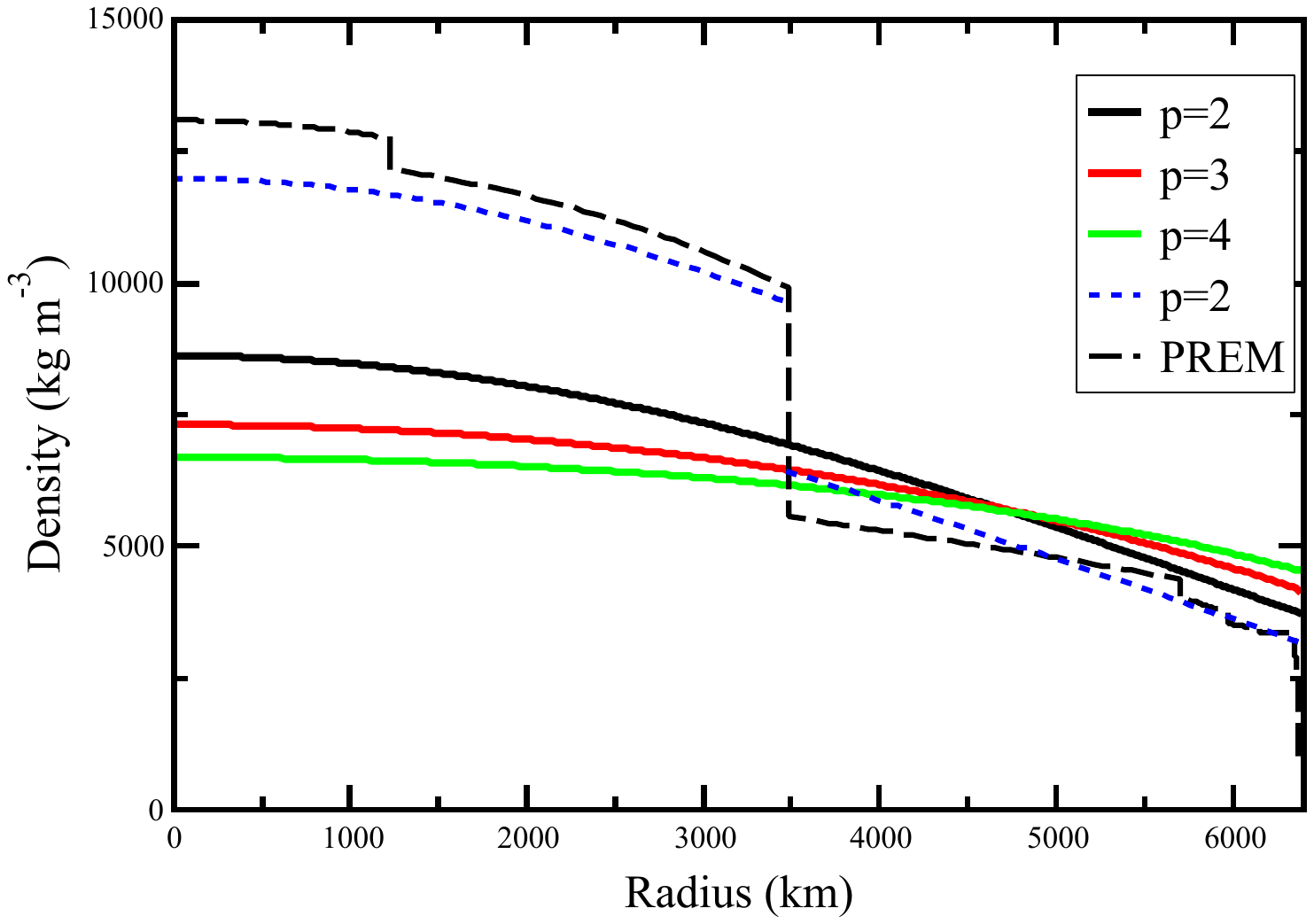}}
\caption{Predicted densities of a Lane-Emden planet with the mass and radius of the Earth, with $p=2$, 3 or 4,  compared to the PREM density model (dotted line).}
\label{earth}
\end{center}
\end{figure}

Figure \ref{earth} shows the density profiles calculated for the Earth for $p=2$, 3 and 4 (black, red, blue, solid lines). The length $r_c$ and density values obtained for various $p$ are listed in Table~\ref{Table2}. The PREM density profile (dotted line) is shown for comparison. Of course, the large discontinuity between core and mantle is not reproduced by our simple model. The prediction with $p=2$ gives an overall better fit to PREM density, although the gradients of curves with $p=3$ or 4 give a better fit to the gradient of PREM, at least in the mantle, that is a better fit to incompressibility. 

The choice of $p=2$ is convenient for checking the accuracy of numerical solutions because the solution is
analytical with $\rho=\rho_c \,\sinc (r/r_c)$ ($\sinc(x)$ is the function $(\sin x)/x$), and $r_c$ is independent of $\rho_c$ ($r_c= r_0=3113$ km, see \eqref{rtilde}). 
However, in addition to the fact that $p=2$ is smaller than experimentally observed, density positivity requires $R\leq \pi r_c=10\,408$ km. Clearly $p=2$ is not appropriate for large exoplanets
like CoRoT-7b. For $p> 2$, the dependence of $r_c$ on $\rho_c$ maintains the positivity of the surface density 
even for very large planets. 

Figure \ref{earth} clearly shows that the question of planet differentiation requires a more complex model. However, the overall behavior of a compressible planet
depends of the volumes subjected to compression, and the volume of Earth's core represents only 16\% of the Earth's volume, so the discrepancy between the actual density and the modeled
density near the center 
is not very significant for the purposes of this paper. Furthermore, if the presence of a metallic core is proven, the construction of a two layers Lane-Emden model is possible. For example, 
in Figure \ref{earth} (blue dashed line), we add a two layer model of the Earth that is in agreement with the Earth's mass, inertia and surface density. We use $p=2$, as the solution is analytical,
but a numerical solution for other values of $p$ would not be difficult. As layered models can only be proposed for a very limited number of objects in the solar system, we will only consider
homogeneous model in the following.

For the different planets considered and for various $p$, we report the length $r_c$ and the planet densities at the surface and at the center in Table~\ref{Table2}. The last column of Table~\ref{Table2} gives the values of the quantity 
\eq{\widetilde M={M\over M_s}={\bar \rho\over \rho_s},}
which is the mass of the planet normalized by the mass $M_s$ of a planet having the same radius but a homogeneous density $\rho_s$
\eq{M_s=\frac{4\pi}{3} \rho_s R^3,\eqlbl{MS}} 
and $\bar \rho$ is the planet's average density.
The value of $\widetilde M$ therefore quantifies the extent to which density is affected by compressibility. As expected, the values of $\widetilde M$ increase with the planet radius.
\begin{table}[htp]
\caption{Quantities computed for various planets and exponents $p$ calculated from our model  of
Lane-Emden planets according to the EoS \eqref{EoS1}. The masses and radii of the planets
are given in Table \ref{Table1}. A }
\begin{center}
\begin{tabular}{cccccc}
\hline
%\\[-0.2 cm]
%$p$    &  $\widetilde R$         & $r_c$     &  $\rho_s$     &  $\rho_c$       & $\widetilde M=M/M_s$ \\
%          &             &    km      & kg m$^{-3}$ & kg m$^{-3}$   & $\,\,\,\,=\bar\rho/\rho_s$ \\
%\hline
%CoRoT-7b &              &              &                      &                 &\\
%2.5 & 2.163 &  4500 &  3780 & 17459 &  2.478 \\
%3 & 1.690 &  5761 &  6009 & 13695 &  1.559 \\
%4 & 1.137 &  8558 &  8092 & 10995 &  1.158 \\
%Earth &              &              &                      &                 &\\
% 2 & 2.046 & 3113 & 3747 & 8625 & 1.471 \\
% 3 & 1.513 & 4210 & 4191 & 7315 & 1.316 \\
% 4 & 1.223 & 5209 & 4549 & 6693 & 1.212 \\
% Mars &              &              &                      &                 &\\
% 2 & 1.089 & 3113 & 3612 & 4438 & 1.089 \\
% 3 & 1.035 & 3274 & 3598 & 4422 & 1.093 \\
% 4 & 0.988 & 3430 & 3583 & 4407 & 1.098 \\
%Mercury &              &              &                      &             &\\
% 2 & 0.784 & 3113 & 5201 & 5774 & 1.043 \\
% 3 & 0.658 & 3709 & 5259 & 5675 & 1.032 \\
% 4 & 0.559 & 4364 & 5303 & 5607 & 1.023 \\
%Moon &              &              &                      &                 &\\
% 2 & 0.558 & 3113 & 3276 & 3452 & 1.021 \\
% 3 & 0.599 & 2900 & 3261 & 3471 & 1.026 \\
% 4 & 0.639 & 2719 & 3243 & 3494 & 1.032 \\
 \\[-0.2 cm]
$p$    &   $r_c$     &  $\rho_s$     &  $\rho_c$       & $\widetilde M=M/M_s$ \\
          &        km      & kg m$^{-3}$ & kg m$^{-3}$   & $\,\,\,\,=\bar\rho/\rho_s$ \\
\hline
CoRoT-7b &              &              &                      &                 &\\
3  &  5761 &  6009 & 13695 &  1.559 \\
4  &  8558 &  8092 & 10995 &  1.158 \\
Earth &              &              &                      &                 &\\
 2  & 3113 & 3747 & 8625 & 1.471 \\
 3   & 4210 & 4191 & 7315 & 1.316 \\
 4   & 5209 & 4549 & 6693 & 1.212 \\
 Mars &              &              &                      &                 &\\
 2  & 3113 & 3612 & 4438 & 1.089 \\
 3   & 3274 & 3598 & 4422 & 1.093 \\
 4   & 3430 & 3583 & 4407 & 1.098 \\
Mercury &              &              &                      &             &\\
2  & 3113 & 5201 & 5774 & 1.043 \\
 3   & 3709 & 5259 & 5675 & 1.032 \\
 4   & 4364 & 5303 & 5607 & 1.023 \\
 Ganymede &              &              &                      &                 &\\
 2  &  3113 &  1848 &  2088 &  1.051 \\
 3 &   2329 &  1737 &  2238 &  1.118\\
4 &   1953 &  1433 &  2509 &  1.355\\
Moon &              &              &                      &                 &\\
2  & 3113 & 3276 & 3452 & 1.021 \\
 3   & 2900 & 3261 & 3471 & 1.026 \\
 4   & 2719 & 3243 & 3494 & 1.032 \\
\end{tabular}
\end{center}
\label{Table2}
\end{table}

\subsection{Mass-radius relations}

\begin{figure*}[!h]
\begin{center}
\vskip -6cm
\centerline{\includegraphics[width=8.8cm]{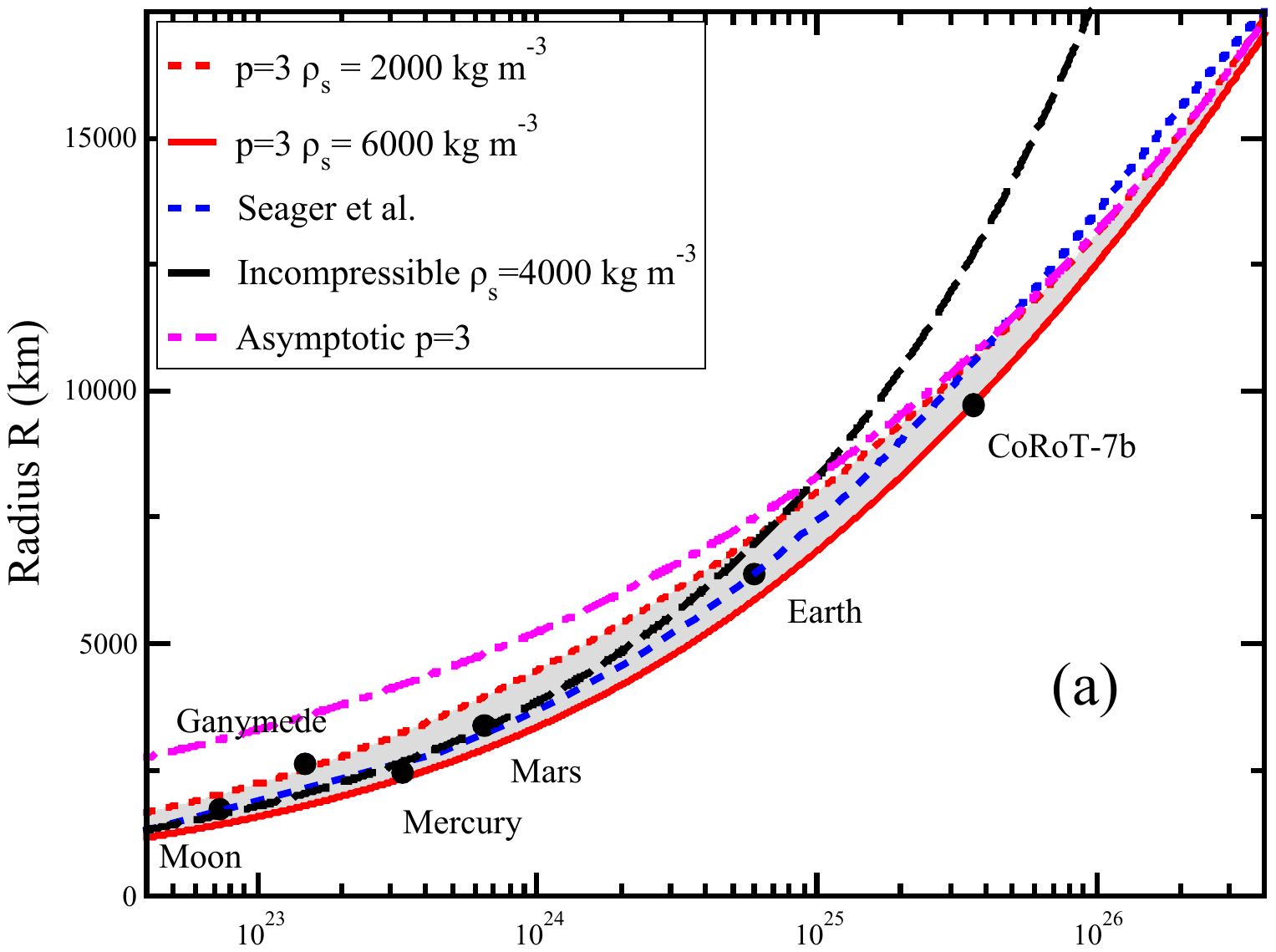}\includegraphics[width=8.8cm]{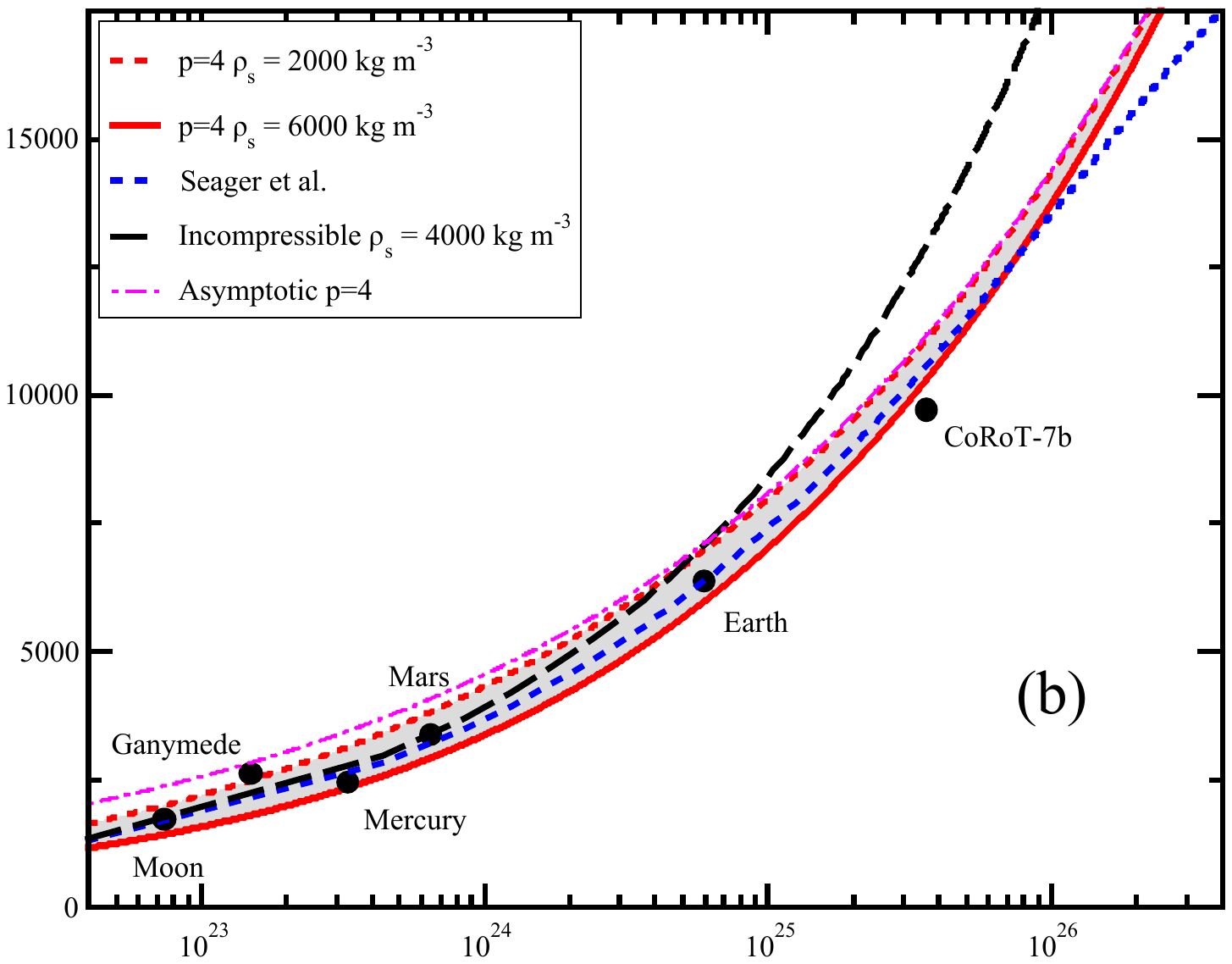}}
\centerline{\includegraphics[width=8.8cm]{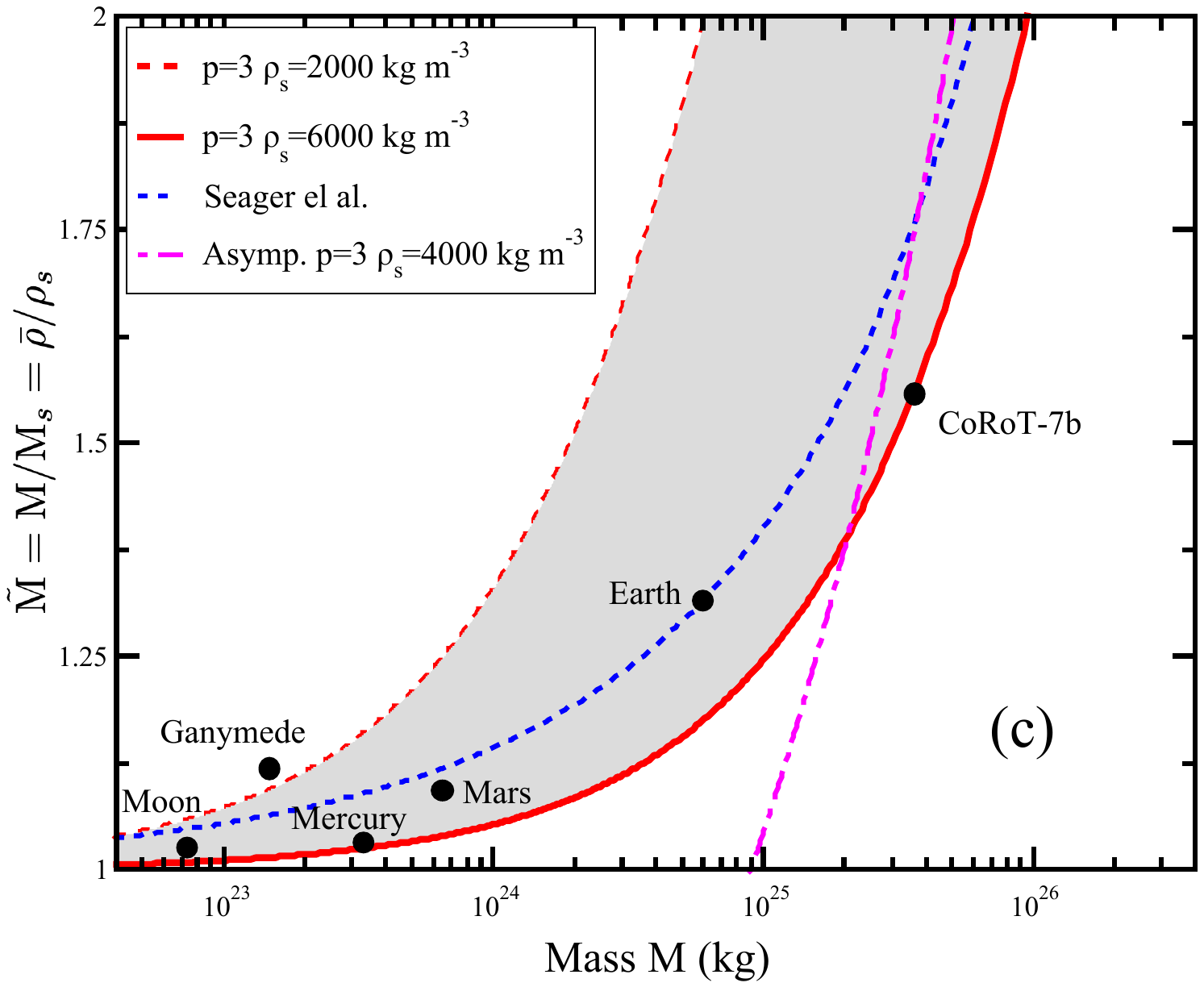}\includegraphics[width=8.8cm]{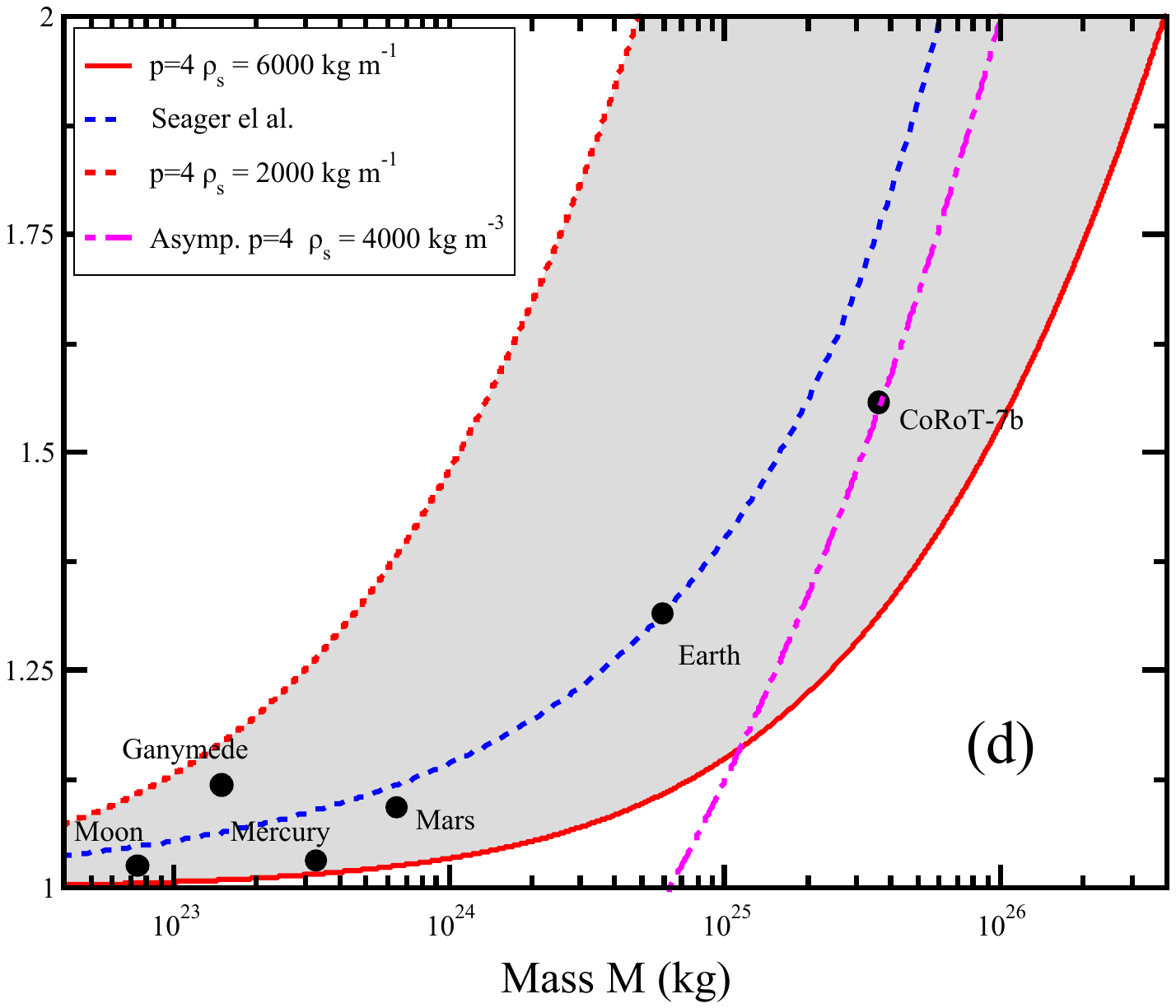}}
\caption{We show the predicted radii for planets of mass 
$M$ and surface densities $\rho_s$, calculated with $p=3$ (panel a) and $p=4$ (panel b). The surface densities
are  2000 kg m$^{-3}$ (red dashed line), or 6000 kg m$^{-3}$ (red solid line). 
In panels c and d, we show the quantity $\widetilde M$ which characterizes the importance of compressibility for each planet. In all panels, the observations for the Moon, Ganymede, Mercury, Mars, Earth and CoRoT-7b are shown with black circles (see values in Tables~\ref{Table1} and \ref{Table2}). 
We also show two limiting cases: the case of an incompressible planet of uniform density 4000 kg m$^{-3}$ (black dashed in panels a and b, and $\widetilde M=1$ in panels c and d) and the asymptotic behavior for very large planets (magenta dashed).
The prediction by \cite{seager07} are shown by blue dashed lines.
}
\label{Figure3}
\end{center}
\end{figure*}

We can also estimate the effect of compressibility according to our model for any planet whose mass is known. Since for a given mass $M$, the radius of the planet depends on its composition, we make two assumptions, one with a rather low surface density $\rho_s=2000$ kg m$^{-3}$, the other with a rather large surface density  
$\rho_s=6000$ kg m$^{-3}$. In Figure \ref{Figure3}, we show the radii $R$ of a planet with mass $M$ and surface density $\rho_s$, in the cases $p=3$ (panel a) and $p=4$ (panel b). They are very similar, with radii slightly larger in the second case, especially for masses $\geq 2\times 10^{25}$ kg. The planet's radius should lie in the shaded area, between the two red curves: closer to the dashed red line for a planet with light composition, and closer to the solid red line  for a planet with a denser composition. For very massives planets, the radius becomes independent of the surface density and the width of the shaded area decreases.
The radii of the planets of Table~\ref{Table1} are also shown in both panels. They lie within or close to the shaded area, closer to the solid line for Mercury and its large core, or for CoRoT-7Bb which probably also has a large core \citep{wagner}, closer to the dashed line for Ganymede rich in water. As indicated in Table~\ref{Table2}, for $p=3$, Ganymede and CoRoT-7b have surface densities slightly outside the 2000-6000 kg m$^{-3}$ interval
of the shaded areas. To make the effect of compressibility more visible,  in Figure \ref{Figure3}c and \ref{Figure3}d, we report the values of $\widetilde M= M/M_s$ as a function of the planet's mass.

The behavior of the predicted mass-radius relationship can be understood with the help of two limiting cases. The gravity of small planets is too low for compressibility
to be an important factor. A thin black dashed line in Figure \ref{Figure3}a shows the radius of an incompressible planet (i.e. $M\propto R^3$), with homogeneous density $\rho_s=4000$ kg m$^{-3}$. 
It lies above the curve corresponding to a compressible planet: compressibility decreases the radius for a given planet mass. An incompressible planet corresponds to $\widetilde M=1$ in
Figure \ref{Figure3}b. Another limiting case is for a very large planet. 
From the mass expression \eqref{mass0} and the expression of $\rho_c$ \eqref{rcKx}, we get that for $p\ne 2$ (for $p=2$ the radius becomes independent of $M$ at large $M$)
\eq{M=4\pi\rho_0 R^3 \left({R\over r_0}\right)^{2\over p-2}  S,}
where
\eq{S={1\over \widetilde R^{3p-4\over p-2}} \int_0^{\widetilde R} \tilde \rho \tilde r^2 \dd  \tilde r.}
When $M\rightarrow +\infty$, $\rho_s/\rho_c\rightarrow 0$ and the normalized radius $\widetilde R$ is bounded by its value when $\tilde \rho=0$ (equal to $\pi$, 1.94 or 1.53 for $p=2$, 3 and 4, see Figure \ref{rc}). The asymptotic value of $S$ can be numerically calculated: $S\rightarrow 0.049$ for $p=3$
and $S\rightarrow 0.144$ for $p=4$. 
%$R= 0.083M^{0.2}$ $M_t=1.04\times 10^{-10}M^{2/5}$}
This result is given here with our notations, but the asymptotic relationship between $M$ and $R$  ($M\propto R^5$ and $M\propto R^4$ for $p=3$ and~4) is a well-known result  \citep[see][p. 98]{chandrasekhar39}. This relationship and that between $\widetilde M$ and $M$ with $p=3$ are represented in Figures \ref{Figure3}a-b by thin magenta dotted-dashed lines. Planets with mass less than 10$^{23}$~kg are basically incompressible, while planets with mass greater than 
10$^{26}$~kg have a radius that follows the asymptotic regime. The Earth has a mass for which none of the limiting cases apply.

Several authors have studied the mass-radius relationship for large condensed planets. They have sometimes considered both a more detailed approach
(with planets including silicate mantles, metallic cores or oceans) and a more sophisticated EoS  (e.g. a third-order Birch–Murnaghan equation). They predictions are however closely similar to ours \citep[see e.g.,][]{valencia,sotin,wagner} with $R\propto M^a$ with an exponent $a$ decreasing with the planet's mass, from 1/3 (incompressible case for small planets)  to an asymptotic value that we found equal to $(p-2)/(3p-4)$, between 0.20 and .25.  For $M_\Earth\leq M \leq 10M_\Earth$ we obtain a value of $\approx$ 0.27, in agreement with previous findings. 

\citet{seager07} uses, like in our paper, a Lane-Emden equation with various possible compositionnal stratifications. 
They propose 
a generic mass-radius relation valid for more or less all compositions, up to a few terrestrial radii, on the form (with our notations)
\eq{\ln\widetilde M\!=\! \left({M\over M_1}\right)^{k_3} \mbox{or   \,\,} R^3={3M\over 4 \pi \rho_s}\exp\!\!\left[- \left({M\over M_1}\right)^{k_3}\right] 
 \eqlbl{seager}}
where they predict an exponent $k_3\approx 0.4$. The equation (23) of \citet{seager07} is written in a more complex form as a relation between $(1/3)\log_{10} (M/\rho_s)$ and $\log_{10} R$, including an adjusted constant $k_1\approx-0.209$ which should be logically $(1/3) \log_{10}(4\pi/3)=-0.207$ to insure a sound behavior when $M\rightarrow 0$. The constant $M_1$ is $m_1(3k_2\ln(10))^{-1/k3}$ using their notations. We choose $M_1= 1.6\times 10^{26}$ kg which corresponds to $m_1=6.5M_\Earth$ which is in the range of values proposed 
in \citet{seager07}, $4.34-10.55M_\Earth$.
Their predictions are also plotted in Figure \ref{Figure3}, panels a and b (blue lines). As their models consider stratified planets in which each building layer verifies a polytropic equation with $p\approx 3-4$, their prediction remains in agreement with our simpler model for the same range of exponent $p$.  However, for very large planet mass, their mass-radius parametrization diverges from the expected asymptotic behavior.

Our approach leads to a mass-radius relationships in perfect agreement with previous, more complex attempts \citep{valencia,sotin,seager07}.
This suggests that the exponent parameter of the Murnaghan EoS (or Lame Emdeen EoS) controls the relationship more than the compositional details.
The advantage of our simple model is that it can easily be perturbed analytically when certain conditions, such as the planet's internal temperature or rate of rotation,
change. This is the subject of the following paragraphs.

\section{Thermal contraction}

What happens now when the temperature of a planet changes while remaining close to an adiabatic state? 
Although these temperature evolutions are small, how is the Earth's radius affected by its cooling rate of 50-100 K Gyr$^{-1}$ \citep{herzberg10}?
The density profile
that we compute does not explicitly include the temperature at each depth. However, from the EoS, the adiabatic temperature profile in the planet can be easily derived when the surface density and therefore the surface temperature is chosen  \citep[e.g.,][]{ricard23}. In this section,
we perturb the surface temperature and calculate the resulting radius change.
This does not change the solution without dimension $\tilde \rho$ of equation \eqref{base}
which is independent of any parameter. However the solution with physical dimensions is affected by changes (denoted with $\delta$) in the quantities $\rho_s$, $\rho_c$, $r_c$ and $u$. Since the mass of the planet does not change, by perturbing \eqref{mass0} we obtain
\begin{multline}
 4\pi \delta \rho_c r_c^3 \int_0^{\widetilde R} \tilde\rho \tilde r^2 \dd\tilde r+
12\pi  \rho_c r_c^2 \delta r_c \int_0^{\widetilde R} \tilde\rho \tilde r^2 \dd \tilde r +\\
4\pi  \rho_c r_c^3 \tilde\rho(\tilde r) \widetilde R^2 \delta \widetilde R=0
\end{multline}
which can be reset using  \eqref{Rsrc}-\eqref{mass0}-\eqref{MS} as
\eq{{\delta \rho_c \over \rho_c} M+
3{\delta r_c \over r_c} M +
3  \left( {\delta R\over R} - {\delta r_c \over  r_c} \right)M_s=0. \eqlbl{a}}
The perturbation of \eqref{rtilde} leads to
\eq{{\delta r_c \over r_c}={p-2 \over 2} {\delta \rho_c \over \rho_c}. \eqlbl{b}}
If the temperature changes while the pressure at the planetary surface is unchanged, the EoS implies that 
the surface density variation is
\eq{{\delta \rho_s\over \rho_s}=-\alpha_s \delta T_s \eqlbl{c},}
where $\alpha_s$ is the thermal expansion at the surface and $T_s$ the adiabatic temperature extrapolated
to the surface (sometimes called the foot of the adiabat).
The definition of $\rho_c$ closes the system.
Indeed, the pertubation of $\rho_s=\rho_c \tilde \rho (\R)$ leads to
\eq{{\delta \rho_s}= \tilde \rho (\R)\delta\rho_c+\rho_c {\dd \tilde \rho \over \dd \tilde r} (\R) \delta \R}
which can be reset as
\eq{{\delta \rho_s\over \rho_s}={\delta\rho_c\over \rho_c}+{\R\over \tilde \rho(\R)} {\dd \tilde \rho \over \dd \tilde r} (\R)  \left( {\delta R\over R} - {\delta  r_c \over r_c} \right).}
However from \eqref{mass} and the definition of gravity $g=GM/R^2$ we obtain
\begin{multline}
{\R\over \tilde \rho(\R)} {\dd \tilde \rho \over \dd \tilde r} (\R)=-{\widetilde M \widetilde R^2\over 3 \tilde\rho^{p-2}(\widetilde R)}=\\
-{\rho_s g R\over K_0} \left( {\rho_0\over \rho_s} \right)^{p}=
-{\rho_s g R\over K_s}, \eqlbl{26}
\end{multline}
where we introduce the surface incompressibility $K_s=K_0 (\rho_s/\rho_0)^p$,
leading to
\eq{{\delta \rho_s\over \rho_s}={\delta\rho_c\over \rho_c}- {\rho_s g R\over K_s} \left( {\delta R\over R} - {\delta  r_c \over r_c} \right). \eqlbl{d}}

Finally using \eqref{a}, \eqref{b}, \eqref{c} and \eqref{d}, we obtain:
\eq{
    {1\over 3} \alpha_s \delta T_s =
    \left( 1+{\rho_s gR\over 3 K_s}{\tilde p \widetilde M\over \tilde p \widetilde M-1} -
     {\tilde p(\widetilde M-1) \over  \tilde p \widetilde M-1}\right) {\delta R\over R},
\eqlbl{resul1}}
with
\eq{\tilde p={p-4/3 \over p-2}.}

The parameter $\tilde p$ is therefore $+\infty$, 5/3 or 4/3 for $p=2$, 3 or 4
(Equation \eqref{resul1} remains valid for  $\tilde p\rightarrow +\infty$ when $p\rightarrow 2$).
In the case of an incompressible fluid, when $K_s\rightarrow \infty$ and $\widetilde M \rightarrow 1$, are radius and temperature simply related by thermal expansivity only. Surprisingly, this is also the case for $\tilde p=0$ when $p=4/3$.

The non-dimensional quantity $\rho_s g R/ K_s$ can also be expressed as the ratio of two quantities well-known to those working in compressible convection, namely
\eq{{\rho_s gR\over K_s}={\mathcal D \over \Gamma},}
where  $\mathcal D$ is the dissipation number and $\Gamma$ the Gr\"uneisen parameter
\eq{\mathcal D= {\alpha_s gR\over C}, \,\,{\mbox{and}}\,\, \Gamma= {\alpha_s K_s \over \rho_s C}.}
In a vigorously convecting planet, $\mathcal D$ controls the slope of the adiabatic temperature and $\mathcal D/\Gamma$ the slope of the adiabatic density \citep[e.g.,][]{ricard23}.

We define as the effective thermal expansion the quantity
\eq{\alpha_e=\alpha_s
    \left( 
    1+{\rho_s gR\over 3K_s} {\tilde p \widetilde M\over \tilde p \widetilde M-1} -
     {\tilde p(\widetilde M-1)\over  \tilde p \widetilde M-1}
     \right)^{-1}.
\eqlbl{effectivealpha}} 
The effective thermal expansion is also
\eq{\alpha_e=\alpha_s { \tilde p \widetilde M-1 \over
    \tilde p -1+{4\pi\rho_s^2 G R^2\over 9 K_s} \tilde p \widetilde M^2 },
} 
where we use $g=(4\pi/3) G \rho_s R \widetilde M$.
When $\widetilde M\rightarrow +\infty$, $\alpha_e\rightarrow 0$ and therefore the effective thermal expansivity is close to 0 for large planets. The effective thermal expansion is
also lower than  $\alpha_e$ when $\widetilde M\approx 1$ if $\tilde p> 1$. In all the cases that we have considered, i.e., $p\geq 2$ which implies $\tilde p > 1$, the compressibility decreases the thermal contraction, $\alpha_e\le \alpha_s$.
However for $0\leq p<2$ which implies $\tilde p \leq 2/3$, $\alpha_e$ becomes larger than $\alpha_s$ when $\widetilde M\approx 1$
and compressibility enhances the contraction of  a small cooling planet. This is why \citet{jaupart15} who assumed a constant incompressibility concluded that compressibility enhances the contraction of the Earth. Their expression is identical to ours when $p=0$ and the compressibility is low. However, using a Murnaghan EoS with $p <2$ (a polytropic index $n\geq 1$) is inappropriate for a  condensed planet and with reasonable exponents $p$, compressibility always decreases the thermal contraction.

\begin{figure}[htbp]
\begin{center}
\centerline{\includegraphics[width=9cm]{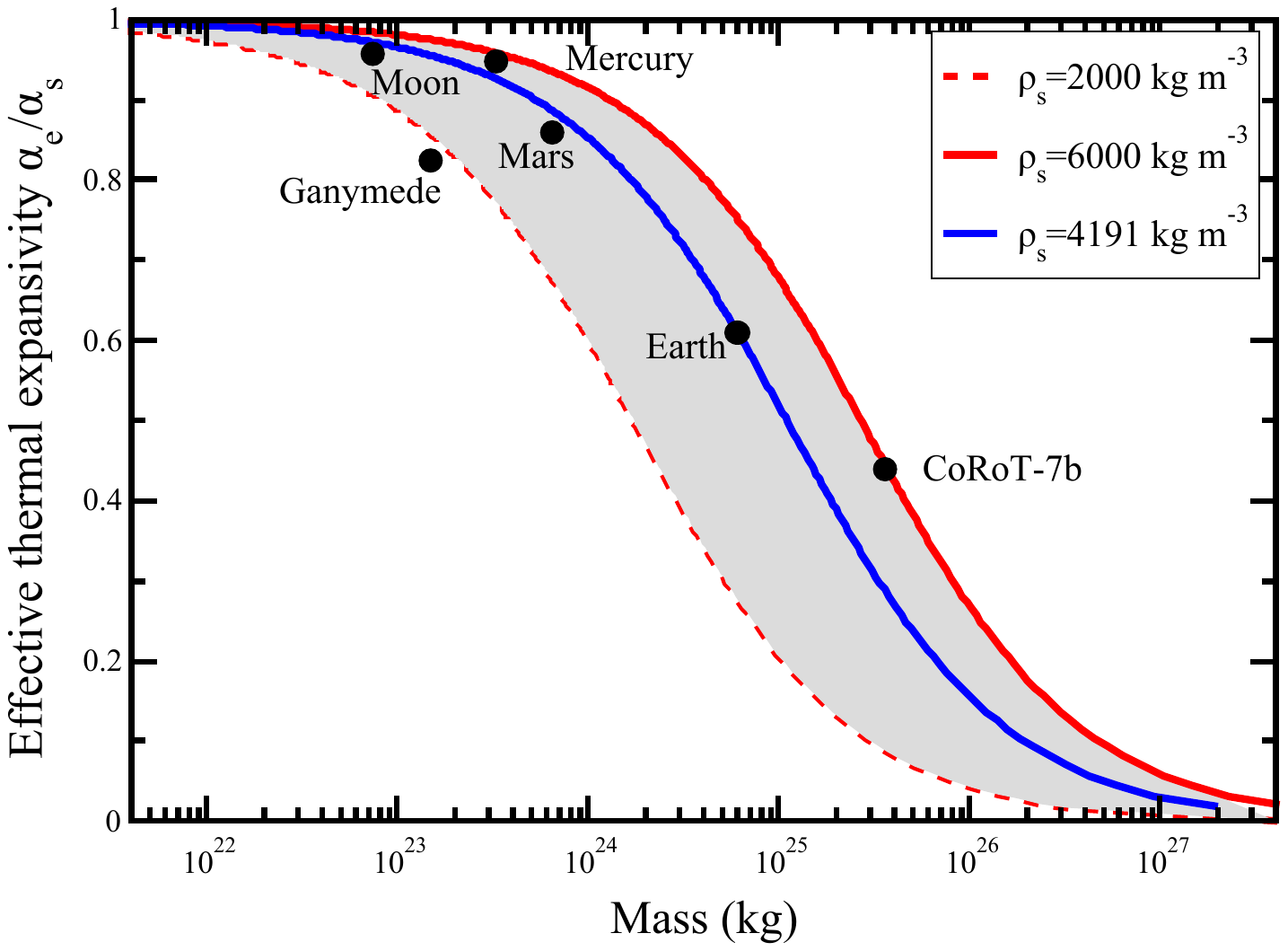}}
\caption{Effective thermal expansivity for $p=3$ of a planet as a function of its mass. We consider three possible surface densities for the condensed
planets: that appropriate for the Earth (solid blue), and for a compositionnally denser (solid red) or lighter (dashed red) planet. Heavy planets like super-Earth,
have a very low effective expansivity as their density is solely related to compressibility.}
\label{Figure4}
\end{center}
\end{figure}

The effective thermal expansion is shown in Figure \ref{Figure4}. We only consider the case $p=3$ and three possible surface densities:
$\rho_s=2000$, $\rho_s= 6000$~kg m$^{-3}$ (same cases as in Figure \ref{Figure3}) and $\rho_s=4191$ kg m$^{-3}$ (the value found for the Earth, see Table~\ref{Table1}). For the various planets we used the values of $\widetilde M$ and $\rho_s$ for $p=3$, from Table \ref{Table1}.

Already for the Earth, the effective
thermal expansivity is significantly reduced compared to its surface value: this is due both to the reduction of $\alpha$ with depth, and to the trade-off between
pressure-temperature-density and gravity. For masses above $10^{26}$ kg the planet's radius becomes insensitive to temperature: the density profile is controlled solely by incompressibility. 

Note that the expression \eqref{resul1} relates the radius to the adiabatic temperature at the surface $T_s$. Using the Eos \eqref{EoS1}, we could easily derive that the adiabatic temperature $T$ and the density are related by \citep[see e.g.,][]{ricard23}
\eq{T=T_s \exp\left[\Gamma \left( {\rho_0\over \rho_s}- {\rho_0\over \rho}\right)\right]. \eqlbl{Tadia}}
The adiabatic temperature increases with depth but this increase remains moderate as
bounded by $\exp(\Gamma)\lesssim 3$ (obtained when $\rho\gg\rho_0$, since  $\rho_s\approx \rho_0$ and $\Gamma\approx 1$). The average temperature $\bar T$ is therefore comparable to $T_s$ in small planets and less than $\approx 3$ times larger
in massive planets. The relative variations of $\delta  T_s/T_s$ and $\delta \bar T/\bar T$ are comparable. If instead of relating the radius changes to the
surface temperature, we rather use the average temperature of the planet, the effective thermal expansion must be further
multiplied by $T_s/\bar T$, $\approx 1$ for small planets,  $\lesssim 1/3$ for large planets.  In this case, $\alpha_e$ decreases even faster with the size of the planet.

\section{Change of rotation velocity}
In this section, we discuss how a change in rotation rate affects the planet's radius.
To account for planetary rotation, a centrifugal potential must be added to Poisson's equation \eqref{poisson}. With this term, equation \eqref{poisson}, verified by density becomes
 \eq{ {1\over r^2} {\dd\over \dd r} \left(K_0 r^2 {\rho^{p-2}\over \rho_0^p} {\dd \rho \over \dd r}\right)+4\pi G\rho-2\Omega^2=0. \eqlbl{dens-with-rot}}
 
Introducing the previously defined  $\tilde \rho$ et $\tilde r$, the equation to solve with rotation, is
\eq{ {1\over \tilde r^2} {\dd\over \dd \tilde r} \left( \tilde r^2 \tilde\rho^{p-2} {\dd \tilde \rho \over \dd \tilde r}\right)+\tilde \rho=\tilde \Omega^2 \tilde \rho(\R) \eqlbl{dens-with-rot-bis}}
where
\eq{\widetilde \Omega^2={\Omega^2\over 2 \pi G \rho_s}.}
In this case, it is not possible to find a generic solution, as previously with $\tilde \rho$, so that the physical solution with dimensions depends only on scaling parameters like $R/r_c$ and $\rho_c/\rho_s$. Here the solution will necessarily depend on a third quantity, $\widetilde\Omega^2$. It therefore seems that, in general, only a numerical solution can be sought if we want to quantify the relationship between the planetary radius and its angular rotation.

Only for $p=2$ the solution can be found analytically and is
\eq{\tilde\rho=\tilde\rho(\R)\left( (1-\widetilde \Omega^2) {\sinc  \,\tilde r\over  \sinc \,\R}+\widetilde \Omega^2\right).}
Using this expression, the mass of the planet is
\begin{multline}
M=4\pi\rho_c r_c^3 \int _0^\R \tilde \rho \tilde r^2 \dd\tilde r=\\
M_s \left( 3(1-\widetilde\Omega^2) F(\R)+ \widetilde \Omega^2 \right) \eqlbl{mass-Omega},
\end{multline}
where
\eq{F(\R)= {\sin \R-\R\cos \R\over \R^2 \sin \R}.}
When the planet's rotation changes, the composition does not change and unlike in the previous section the surface temperature is unaffected. So, by changing $\Omega$, the density will change at depth but not the boundary conditions at the surface that maintain $\rho_s$ fixed.

Therefore, equation \eqref{mass-Omega} relates the rotation rate of a planet to its radius which appears both in $\R=R/r_c$ and in $M_s=4\pi \rho_s R^3/3$.
We can now differentiate \eqref{mass-Omega} taking into account that $\delta M=0$, $\delta M_s=3M_s \delta R/R$, $\delta F=(\dd F/\dd \R)\delta \R$, $\delta \R=\R \delta R/R$, and get
\begin{multline}
3\left( (1-\widetilde\Omega^2) (3F+ \R{\dd F\over \dd \R}) + \widetilde\Omega^2  \right){\delta R\over R}=\\
2 \widetilde\Omega^2(3 F-1) {\delta \widetilde\Omega\over  \widetilde\Omega}.
\end{multline}
This expression can be simplified through a rather cumbersome algebra. First the definition of $F$ leads to
\eq{3F+ \R{\dd F\over \dd \R}=1+\R^2 F^2,}
where $F$, extracted from the mass conservation \eqref{mass-Omega} is
\eq{F={\widetilde M- \widetilde\Omega^2\over 3 (1-\widetilde\Omega^2)} }
with  again $\widetilde M=M/M_s$.
Defining the parameter 
\eq{m={\Omega^2 R\over g},} 
which is central in the planet's hydrostatic theory,
allows us to write $\widetilde\Omega^2$ as
\eq{\widetilde\Omega^2={2\over 3}m\widetilde M.}
Using \eqref{26} and all simplifications done, we obtain

\eq{{\delta R\over R}={4\over 9}  {m \widetilde M (\widetilde M-1) \over 1 -{2\over 3} m \widetilde M+{ \rho_s g R\over 3 K_s } \widetilde M {(1- {2\over 3} m )^2 } }
{\delta \Omega\over \Omega}.
}
The $\widetilde M-1$ term insures that the planetary radius is independent of the rotation rate when the planet is incompressible.

By comparison with the result of \cite{saito74} (see Eq. \ref{eq:saito}), our model predicts a Love number
\begin{multline}
h_0^*  ={2 \widetilde M (\widetilde M-1) \over 3 -{2} m \widetilde M+{ \rho_s g R\over K_s } \widetilde M {(1- {2\over 3} m )^2 }}\\
\approx { 2\widetilde M (\widetilde M-1) \over 3 +{ \rho_s g R\over  K_s } \widetilde M } \label{eq:loveresult}
\end{multline}
where the terms in $m$ can generally be omitted ($m=1/289$ for the Earth). The parameter $ \rho_s g R / K_s= \mathcal{D}/\Gamma $ that we discussed in the previous section appears again.

\begin{figure}[htbp]
\begin{center}
\centerline{\includegraphics[width=8.6cm,angle=0]{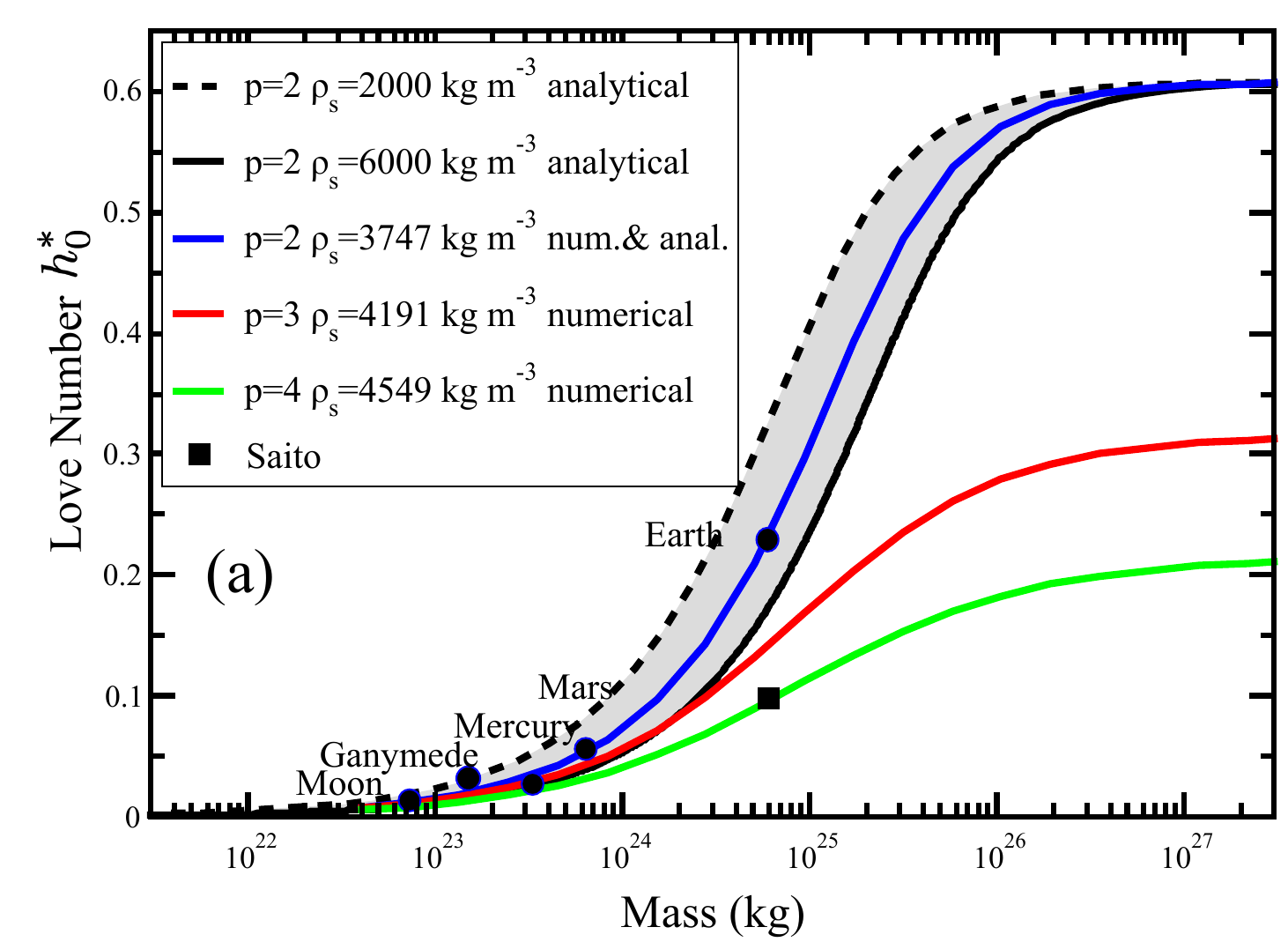}}
\centerline{\includegraphics[width=8.6cm,angle=0]{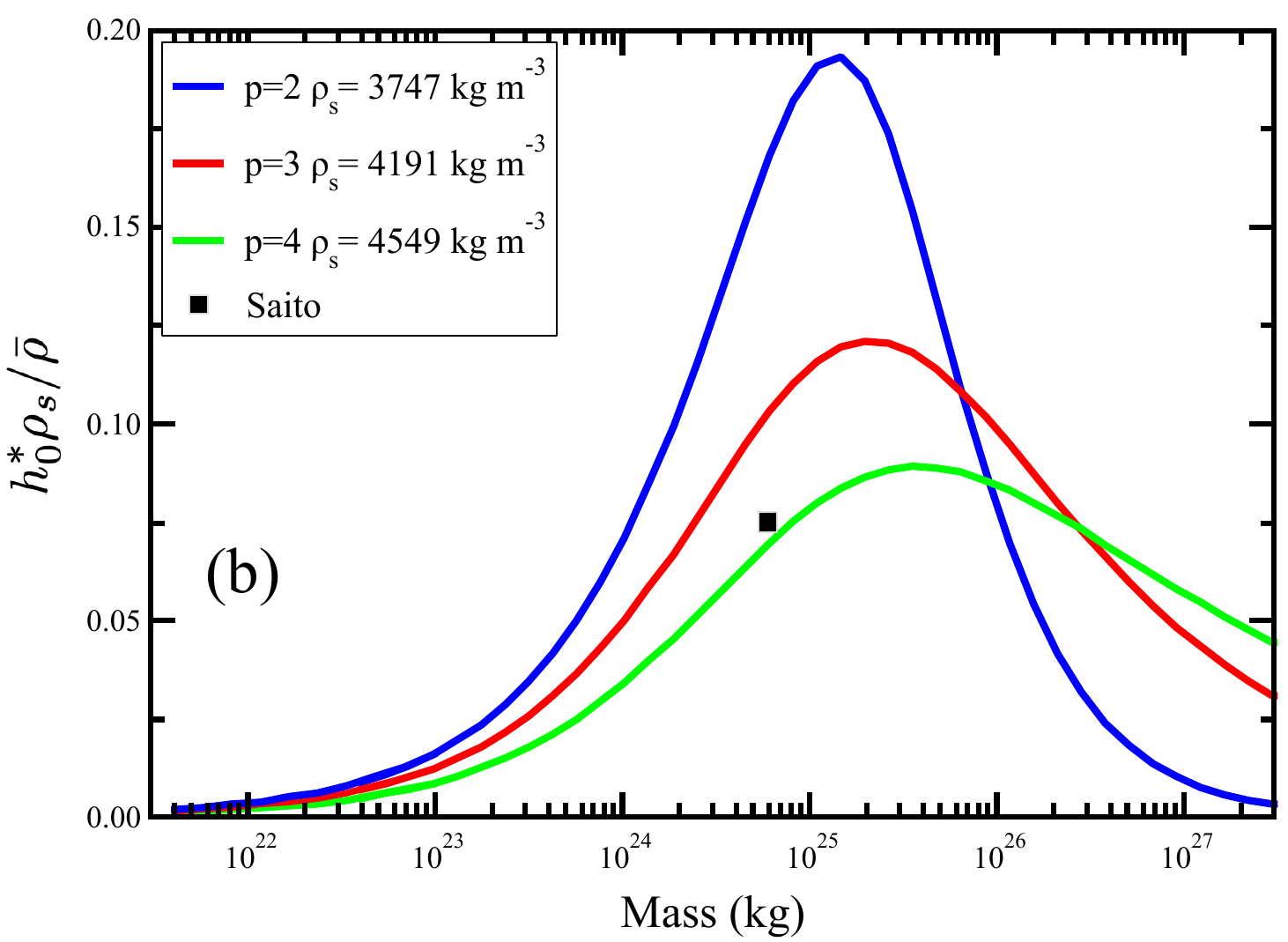}}
\caption{
a) Degree 0 rotational Love number as a function of the planetary mass. The curves depict the analytical solution for $p=2$ (Eq. \ref{eq:loveresult}) for three values of the surface density. The filled circles correspond to the analytical solution for various planets, using the values of the surface density taken from Table~\ref{Table1} with $p=2$ (we excluded CoRoT-7b, too massive for the case $p=2$). The red and green curves are the numerical solution of the equations for $p=3$ and 4, and the black square is Saito's value. b) As the contraction of a planet during despinning contains the flattening term $m$ which decreases as $1/\bar \rho$, we plot $h^*_0 \rho_s/\bar\rho$ which accounts all the terms depending of the planet's mass, as a function of the planet's mass. The prediction of Saito corresponds to a value of
$p$ slightly lower than 4. 
}
\label{Figure5}
\end{center}
\end{figure}

We plot in  Figure \ref{Figure5}a, the Love number $h^*_0$ as a function of planetary mass for a planet having three possible surface densities $\rho_s=2000$, $\rho_s=$ 6000 kg m$^{-3}$ (same cases as in Figure \ref{Figure3} and Figure~\ref{Figure4}) and $\rho_s=3747$ kg m$^{-3}$ (the value found for the Earth, see Table~\ref{Table1}) and $m\ll 1$. When the mass is large, $h^*_0$ reaches an asymptotic value, but in this case $m=\Omega^2 R/g= 3\Omega^2/ (4\pi G\bar \rho)$ goes to zero
and the radius becomes independent of the rotation rate (see \eqref{saito}). This is more conspicuous in Figure \ref{Figure5}b, where we plot $h^*_0 \rho_s/\bar \rho$ that includes all the terms dependent of the mass. Small planets are incompressible and their
radius is independent of the rotation rate, large planets have such a large density that $h^*_0 \rho_0/\bar \rho $  is very small, the Earth and CoRoT-7b are in a mass range where their radius is most sensitive to their rotation rate.

The value we obtain for the Earth, $h^*_0=0.23$ is twice as large as the value proposed by \citet{saito74}. Our analytical model, however uses $p=2$ which is too small a value. To confirm that this large $h^*_0$ is only due to an inappropriate
choice of $p$, we take a brute-force approach and numerically solve \eqref{dens-with-rot} for $p=2$, 3 or 4, and for a surface density of 3747, 4191 and 4549 kg m$^3$,  respectively (the values of surface densities listed in Table \ref{Table1} for the Earth). The rotation rate is set successively to
the rotation period of the Earth $\Omega_e$, then to $\Omega_e-\delta \Omega_e$ and to $\Omega_e+\delta \Omega_e$. Noting the corresponding radii $R^{-}$ and $R^{+}$, the Love number is approximated by (see Eq. \ref{eq:saito})
\eq{h_0^*={3\over 4mR}{R^+-R^- \over \delta \Omega_e}.}

The results of these numerical experiments are also shown in Figure \ref{Figure5}a. The numerical estimate for $p=2$, in perfect agreement with the analytical result, is not repeated.
For $p=3$, the numerical estimate for the Earth is closer to the value proposed by \citet{saito74}  and for $p=4$, it becomes
basically identical (even slightly lower). This confirms that a $p=2$ model is too compressible and leads to a planet that is excessively sensitive to changes in temperature or rotation speed.

\section{Discussion and Conclusion}

The Lane-Emdden models have been widely used in astrophysics. Their application to condensed planets has been less so
although the EoS of Murnaghan's EoS \citep{murnaghan37}, often used in geophysics for silicate ou metal planets, belongs
to the family of polytropic EoS. An important difference between astrophysical models of stars or gas planets and telluric planets comes from a difference in the polytropic exponent $n$. Another difference is that the ratio between central and surface densities in telluric planets is never very large so that the surface boundary conditions remain crucial while for say, a giant gas planet, the 
surface density is zero and the density profile or the mass only depend on the central density $\rho_c$. 

For the Earth, an accurate elastic model of compressibility is known and even for several objects of the solar system, more realistic compressibility than Lane-Emden models can be
proposed. The aim of this article is to propose a simple generic model for all planets whose properties are not precisely known. For any specific planet, a more detailed
model could account for their composition, differentiation and phase changes. At any rate, for the smallest objects (Moon, Mercury and Mars), the assumption of a total incompressibility makes little difference: the change in radius
is mainly linked to thermal expansivity and is insensitive to rotation. For the Earth compressibility plays a minor but significant role (see
Figures \ref{Figure4} and \ref{Figure5}), for CoRoT-7B or larger super Earth, it becomes a major effect.

In this paper, we have refrained from making numerical applications of radius changes for the planets that we have previously examined (Earth, Mars, Mercury and the Moon); there are not very different from previous estimates. The
radius decrease due to the cooling of the Earth (say 250 K in 3 Gyr with $\alpha_0=3\times 10^{-5}$~K$^{-1}$) should be around 9 km using $\alpha_e/\alpha_s=0.6$ in Figure \ref{Figure4}.  The Earth's sideral day was only 13 or 15 hours, during the Archean, 3.2 Gyr ago, see e.g. \citep{farhat22} or \citep{eulenfeld23}, using the geological records of tidalites \citep{eriksson01}. This implies a further radius reduction of 1.4 km 
 (as $\Omega$ varied significantly we integrate \eqref{saito} and use $\delta R=(1/3)h_0^*m R \left(1-(\Omega_A^2/\Omega_e^2)\right)$
 where $\Omega_A$ is the Archean rotation rate).
 Of course, plate tectonics has erased all evidence of these contractions which only reach $\approx 3~\mu$m~yr$^{-1}$; only on planets whose lithosphere has been frozen for billions
 years can thermal contractions be observed. 

Our main objective  was to show that the important parameters controlling the changes of radius are the dissipation number $\mathcal{D}=\alpha_s g R/C$ and the Gr\"uneisen parameter $\Gamma=\alpha_s K_s/(\rho_s C)$. The Gr\"uneisen parameter varies little between 1 and 2 for most planet's compositions. On the contrary, very large dissipation numbers are specific to super-Earths since $R$ and $g$ increase
with the planet's mass \citep{ricard23}. From for Figures \ref{Figure4} and \ref{Figure5}, it is clear that Earth lies in something of a transition zone, between
smaller objects affected by temperature, and larger objects which are mostly insensitive to temperature but affected by their angular rotation. For the large exoplanets that have been discovered, dissipation numbers larger than 10 are expected ($\mathcal{D}\approx 0.6$ for the Earth) from the observed radius and masses \citep[see e.g.,][]{otegi20}. 
In the range $10^{24}-10^{26}$ kg and for planets those interiors are largely unknown, our approach can provide first order estimates of density profiles and potential changes of the planetary radii through time.
\\
\\
Acknowledgements: this research was founded by the French National Program of Planetology (PNP, CNRS-INSU, Proposal: Compressibility and Convection in Condensed Planets ). 

\bibliographystyle{aasjournal}
\bibliography{biblio_Lane-Emden}

%
%%% This command is needed to show the entire author+affiliation list when
%%% the collaboration and author truncation commands are used.  It has to
%%% go at the end of the manuscript.
%%\allauthors
%
%%% Include this line if you are using the \added, \replaced, \deleted
%%% commands to see a summary list of all changes at the end of the article.
%%\listofchanges

\end{document}